\documentclass[final,3p,times,authoryear]{elsarticle}

\usepackage{amsmath}
\usepackage{amsfonts}
\usepackage{amssymb}
\usepackage[titletoc,title]{appendix}
\usepackage{booktabs}
\usepackage[mmddyyyy]{datetime}
\usepackage{dutchcal}
\usepackage{enumitem}
\usepackage{graphicx}
\usepackage{hyperref}
\usepackage[utf8]{inputenc}
\usepackage{lineno}
\usepackage[percent]{overpic}
\usepackage{pgfplots}
\usepackage{relsize}
\usepackage{subfigure}
\usepackage{siunitx}
\usepackage{tikz}

\modulolinenumbers[5]
\numberwithin{equation}{section}

\usetikzlibrary{calc}
\usetikzlibrary{plotmarks}
\usetikzlibrary{positioning}
\usetikzlibrary{fit}
\usetikzlibrary{decorations.pathmorphing}
\usetikzlibrary{backgrounds}
\pgfplotsset{compat = newest}
\pgfplotsset{ legend style={font=\tiny} }
\definecolor{bgreen}{rgb}{0.0,0.5,0.0}
\definecolor{bblue}{rgb}{0.0,0.0,0.9}
\definecolor{bgold}{rgb}{0.7,0.5,0.0}
\definecolor{bred}{rgb}{0.9,0.0,0.0}

\begin{document}
	
	\begin{frontmatter}
		
		
		
		\title{Origin of the power-law profile in a core-collapsing galactic globular-cluster model}
		
		
		\author{Yuta Ito}
		\ead{yutaito30@gmail.com}

		\begin{abstract}
			Observed galactic globular clusters reveal power-law structural profiles in the inner halos around the core-collapse stage. However, the origin of the power-law has not been explained in an acceptable manner. The present paper applies the Buckingham's Pi theorem to the orbit-averaged Fokker-Plank (OAFP) model of equal masses to study the inner-halo structure of a core-collapsing isotropic star cluster. We first prove that an infinite OAFP model evolves self-similarly because of the principle of covariance. We then show that the inner halo must form a power-law profile in a finite OAFP model that has complete similarity so that the principle of covariance and conservation laws hold. The conventional assumption that inner halos are self-similar and stationary is unnecessary to explain the power-law profiles.
		\end{abstract}
		
		\begin{keyword}
			Star clusters\sep Globular Clusters\sep Nuclear star clusters\sep Relaxation evolution\sep  Collisional stellar dynamics\sep Kinetic theory \sep Self-similar evolution \sep Dimensional Analysis \sep Buckingham's Pi theorem \sep Re-normalization group 
		\end{keyword}
		
	\end{frontmatter}
	
	\section{Introduction}

Globular clusters are observed ubiquitously in galaxies. Our Galaxy hosts approximately 160 clusters \citep{Harris_1996, Harris_2010}. One of the most unique features is that about 20\% of Galactic globular clusters reveal steep power-law structural profiles in the halos surrounding high-density cores \citep{Djorgovski_1986}. According to the standard stellar-dynamics theory \citep{Spitzer_1988, Heggie_2003, Binney_2011}, the high-density cores are the result of ``the gravothermal instability \citep{Lynden_Bell_1968, Lynden_Bell_1980}'' caused by the negative specific heat originating from slow-relaxation and self-gravity. As heat flows from the dense core to the sparse halo through two-body relaxation, the core gets hotter and hotter. During the heat release, high-energy stars leave the core and low-energy stars falls in a deeper self-consistent core potential, making the core smaller and, in a relative manner, the surrounding halo larger and more elongated. This instability is considered to last until the core completely collapses. The gravothermal instability itself, however, can not explain why observed halos could have power-law profiles. 

The power-law halo structures have been conventionally explained by assuming self-similarity. The simplest self-consistent model for galactic globular clusters is the orbit-averaged Fokker-Planck (OAFP) model \citep{Henon_1961, Cohn_1979, Cohn_1980}. \citet{Henon_1961} pioneered and assumed the infinite OAFP model to evolve self-similarly, showing that the self-similar state is stable against a linear perturbation, while the Boltzmann-Maxwell distribution function is not. \citet{Cohn_1980} confirmed the self-similarity in the core and inner halo, by numerically solving the fully time-dependent OAFP model in a tidal field. \citet{Lynden_Bell_1980} provided a possible physical reasoning of the power-law profile formation. After an inner halo receives heat from the shrinking high-density core, it is less active than the core on the time-scale of heat flow and can be considered a stationary remnant. In this case, the core-halo profile could always have the same structure but on different spatial scales. Accordingly, the power-law was introduced as a result of self-similarity and stationary state. Later, \citep{Heggie_1988} revisited the self-similar OAFP model, assuming that the inner halo is a power-law profile, and reproduced \citet{Cohn_1980}'s result. 

The previous studies lack explaining (i) why local physical phenomena are enough to explain the large structures of globular clusters, without using a global analysis, and (ii) to what extent 'ideal' self-similarity can be used to explain physical phenomena without specifying the kind of similarity that globular clusters could have. The assumptions of self-similarity and stationary state, based on the local analysis, rely on the authors' intuitions or numerical results only. Especially, the self-similarity assumption does not apply to realistic finite globular clusters. Self-similarity is an ideal characteristic and observed generally in an infinite physical system or a system with moving boundary conditions. To figure out the cause of approximate or intermediate self-similarity in a realistic finite system, dimensional analysis must be used. 

Dimensional analysis studies how to (i) systematically transform dimensional physical quantities to dimensionless for numerical calculation or mathematical analysis and (ii) minimize the number of the dimensionless quantities \citep{Sedov_1959, Barenblatt_2005}. The minimization process is carried out by the main concept, Buckingham's Pi-theorem, and the re-normalization group method. The latter is well-known for various scientific communities as a tool to find the self-similarity of time-dependent systems. \citep{Henon_1961} first applied the method to the OAFP model. On the one hand, the Buckingham's Pi theorem relies on the principle of covariance associated with physical dimensions and units. It has been a fundamental tool to discuss a broad range of self-similarity and predict a possible time-evolution of physical quantities. However, the theorem has been studied mainly by applied mathematicians and not been the primary tool for physicists and astronomers. In stellar-dynamics studies, the theorem was used in only few works for collisionless dynamics \citep{Henriksen_1995, Henriksen_1997, Henriksen_2015}, but not for collisional dynamics. 

The present paper revisits a classical problem in galactic globular-cluster studies. By applying the Buckingham's Pi theorem to an infinite and a finite isotropic OAFP model of equal masses, it shows that the principle of covariance and conservation laws can sensibly explain the formation of power-law profiles the inner halos. It is organized as follows. Section \ref{sec2:Dim_Analy} explains the Buckingham's Pi theorem and highlights conditions used for dimensional analysis. Sections \ref{sec3:Inf_OAFP} and \ref{sec4:Fin_OAFP} apply the theorem to the infinite and finite OAFP models. Section \ref{sec5:Conclusion} is Conclusion.   
    
 	\section{Buckingham's Pi theorem, kinds of similarity, and conditions used for dimensional analysis}\label{sec2:Dim_Analy}

\subsection{Buckingham's Pi theorem}
Buckingham's Pi theorem is the kernel of dimensional analysis and a mathematical tool to convert a set of physical quantities into dimensionless forms in a physical system \citep{Sedov_1959,Barenblatt_2005}. Assume that the physical system is completely described by a physical quantity $a$ in terms of governing parameters (physical quantities) 
\begin{equation}
	a=w\,\left(a_{1}, a_{2}, \cdots, a_{k},b_{1}, b_{2}, \cdots, b_{m}\right),\label{eq.a}
\end{equation}
where $w$ is a physical relation (or a function) of $\{a_{1}, a_{2}, \cdots, a_{k}\}$ $(k=1,2,\cdots)$ that have independent dimensions and $\{b_{1}, b_{2}, \cdots, b_{m}\}$ $(m=1,2,\cdots)$ that have dimensions dependent on $\{a_{1}, a_{2}, \cdots, a_{k}\}$, as follows,
\begin{eqnarray}
	\left[b_{1}\right]&=&\left[a_{1}\right]^{p_{1}}\left[a_{2}\right]^{q_{1}}\cdots\left[a_{k}\right]^{r_{1}},\\
	\left[b_{2}\right]&=&\left[a_{1}\right]^{p_{2}}\left[a_{2}\right]^{q_{2}}\cdots\left[a_{k}\right]^{r_{2}},\\
	     &\,&\qquad\vdots\nonumber\\
	\left[b_{m}\right]&=&\left[a_{1}\right]^{p_{m}}\left[a_{2}\right]^{q_{m}}\cdots\left[a_{k}\right]^{r_{m}},
\end{eqnarray}
where $\left[a_{l}\right]$ means the dimension of physical quantity $a_{l}$, and $\{p_{1}, p_{2}, \cdots, p_{m}\}$, $\{q_{1}, q_{2}, \cdots, q_{m}\}$, $\cdots$, $\{r_{1}, r_{2}, \cdots, r_{m}\}$ are sets of real numbers. If we introduce dimensionless parameters
\begin{equation}
	\Pi\equiv\frac{a}{a_{1}^{p_0}a_{2}^{q_0}\cdots a_{k}^{r_0}},\quad
	\Pi_{1}\equiv\frac{b_{1}}{a_{1}^{p_1}a_{2}^{q_1}\cdots a_{k}^{r_1}},\quad
	\Pi_{2}\equiv\frac{b_{2}}{a_{1}^{p_2}a_{2}^{q_2}\cdots a_{k}^{r_2}},\quad
	\cdots,\quad
	\Pi_{m}\equiv\frac{b_{m}}{a_{1}^{p_m}a_{2}^{q_m}\cdots a_{k}^{r_m}},
\end{equation}
where $\{p_0, q_{0},\cdots,r_{0}\}$ is a set of real numbers, then equation \eqref{eq.a} reduces to
\begin{equation}
	\Pi=W\,\left(a_{1}, a_{2}, \cdots, a_{k},\Pi_{1}, \Pi_{2}, \cdots, \Pi_{m}\right),\label{eq.Pi} 
\end{equation}
where $W$ is a physical relation. Since physical relations, including laws of physics, should not depend of the choice of physical units (\emph{the principle of covariance}), $\Pi$ and $\{\Pi_{1},\Pi_{2}\cdots\Pi_{m}\}$ must be independent of changes in the governing parameters $\{a_{1}, a_{2}\cdots a_{k}\}$. Equation \eqref{eq.Pi} then reads
\begin{equation}
	\Pi=\Omega\,\left(\Pi_{1}, \Pi_{2}, \cdots, \Pi_{m}\right),\label{eq.Pi_f} 
\end{equation}		
where $\Omega$ is a physical relation. Based on the arguments above, \emph{the Buckingham's Pi theorem} states the algebraic relationship that the number of dimensionless governing parameters $\{\Pi_{1}, \Pi_{2}, \cdots, \Pi_{m}\}$ equals the total number $n$ of the dimensional governing parameters $\{a_{1}, a_{2}, \cdots, a_{k},b_{1}, b_{2}, \cdots, b_{m}\}$ minus the number of the governing parameters with independent dimensions $\{a_{1}, a_{2}, \cdots, a_{k}\}$, that is,
\begin{equation}
	m=n-k.
\end{equation}

\subsection{Kinds of similarity}\label{subsec:Kind_ss}

When one of the dimensionless governing parameters $\{\Pi_{1}, \Pi_{2}, \cdots, \Pi_{m}\}$ is very small or large, say $\Pi_{m}\to0$ or $\Pi_{m}\to\infty$,  there could be three possible resulting cases \citep{Barenblatt_2005}. The first case is the complete similarity, or the similarity of the first kind. As $\Pi_{m}\to0$ or $\Pi_{m}\to\infty$, $\Omega$ is finite and nonzero
\begin{equation}
	\Pi\approx\Omega\,\left(\Pi_{1}, \Pi_{2}, \cdots, \Pi_{m-1}\right).
\end{equation} 
In this case all the similarity variables are determined within dimensional analysis. The second case is the incomplete similarity, or the similarity of the second kind. As $\Pi_{m}\to0$ or $\Pi_{m}\to\infty$, $\Omega$ approaches zero or infinity. However, the intermediate asymptotic behavior of $\Omega$ still can be described by a reduced number of arguments 
\begin{equation}
	\Pi\approx\Pi_{m}^{\beta}\,\Omega\,\left(\frac{\Pi_{1}}{\Pi_{m}^{\beta_1}}, \frac{\Pi_{2}}{\Pi_{m}^{\beta_2}}, \cdots, \frac{\Pi_{m-1}}{\Pi_{m}^{\beta_{m-1}}}\right),
\end{equation} 
where $\{\beta, \beta_1,\beta_2,\cdots,\beta_{m-1}\}$ are real numbers. In this case, these power indexes can not be determined within dimensional analysis, as $b_m$ is the essential factor to determine the asymptotic behavior of $\Omega$. The third case is that, as $\Pi_{m}\to0$ or $\Pi_{m}\to\infty$, $\Omega$ approaches zero or infinity and does not show any features of similarity.

\subsection{Conditions used in the present paper for dimensional analysis}\label{subsec:Conditions}

In general, applications of the Buckingham's Pi theorem are not simple tasks. It could provide mathematically/physically both correct and incorrect results. The results must be always compared to physical arguments and the numerical results of fully time-dependent models. Yet, the following conditions simplify our arguments in applying the Buckingham's Pi theorem to the OAFP model.
\begin{itemize}
	\item Condition 1: Our interest is only a full or approximate self-similar phenomenon that is comparable to the numerical results of the OAFP model obtained in \cite{Cohn_1980,Heggie_1988}.
	\item Condition 2: Stars with very high specific energies are collision-less and do not contribute to the relaxation processes.
	\item Condition 3: The physical quantities to characterize the OAFP model monotonically decrease or increase with the cluster radius.
\end{itemize}
Condition 1 means that we focus on the relation of self-similarity with the power-law profile found in the previous results. In the fully time-dependent finite OAFP model \citep{Cohn_1980}, the normalized density and probability distribution function do not approach neither zero nor infinite with time. The corresponding self-similar model \citep{Heggie_1988} reproduced the results as an eigenvalue problem. These results implicate that the finite OAFP model have complete similarity in the limits of proper dimensionless quantities. We exclude any incomplete similar and non-similar phenomena from our dimensional analysis. Condition 2 is a basic property of the outer halo in the finite OAFP model. This condition excludes the region, where dimensional analysis is not applicable, from our discussion. As explained in Section \ref{subsec:Kind_ss}, similarity analysis expects the physical relation $\Omega$ to be non-zero. If $\Omega$ is zero, it is singular. The product $\Pi_{m}^{\beta}\,\Omega$ is mathematically ill-defined and undetermined. Condition 3 is based on the Virial equilibrium of an isotropic star cluster on dynamical-time scales \citep{Spitzer_1988,Binney_2011}. On this scale, if the cluster's distribution function monotonically decreases with specific energy, it is stable. We may expect this stability to hold on the relaxation-time scales during the core-collapse. 

 	\section{Dimensional Analysis on an infinite OAFP model}\label{sec3:Inf_OAFP}

\citet{Henon_1961} introduced the OAFP model for an isotropic star cluster of $N$ equal stellar masses $m$. In the cluster, stars can have energy $\epsilon$ per unit mass.  The model is described by three physical quantities, the phase-space probability density $f(\epsilon,t)$ of finding a star with specific energy $\epsilon$ at time $t$, the self-consistent Newtonian gravitational potential $\phi(r,t)$ at cluster radius $r$, and $q(\epsilon,t)$ that is proportional to the phase-space volume occupied by stars with a specific energy below or equal to $\epsilon$. The time-evolution of these quantities obeys the OAFP equation, Poisson equation, and $q$-integral (the definition of $q(\epsilon,t)$)
\begin{eqnarray}
	0&=&\frac{\partial\, f}{\partial\, t}\frac{\partial\, q}{\partial\, \epsilon}-\frac{\partial\, f}{\partial\, \epsilon}\frac{\partial\, q}{\partial\, t}-16\pi^{2}G^{2}m^2N\ln[\Lambda]\nonumber\\
	&\,&\quad\times\frac{\partial\, }{\partial\, \epsilon}\left[f(\epsilon,t)\int^{\epsilon}_{\phi(0,t)}f\left(\epsilon',t\right)\frac{\partial\, q}{\partial\, \epsilon’}\text{d}\epsilon'+\frac{\partial\, f}{\partial\, \epsilon}\left(\int^{\epsilon}_{\phi(0,t)}f(\epsilon',t)q(\epsilon',t)\text{d}\epsilon'+q(\epsilon,t)\int_{\epsilon}^{0}f(\epsilon',t)\text{d}\epsilon'\right)\right],\\
	0&=&r(\phi,t)\,\frac{\partial^{2}\, r}{\partial\, \phi^2}-2\left(\frac{\partial\, r}{\partial\, \phi}\right)^{2}+16\pi^{2}GMN\,r(\phi,t)\left(\frac{\partial\, r}{\partial\, \phi}\right)^{3}\int^{0}_{\phi}\sqrt{2(\epsilon'-\phi)}f(\epsilon',t)\text{d}\epsilon',\\
	q(\epsilon,t)&=&\frac{1}{3}\int_{\phi(0,t)}^{\epsilon}\left[2(\epsilon-\phi')\right]^{3/2}\,r(\phi',t)^{2}\frac{\partial\, r}{\partial\, \phi'}\text{d}\phi',
\end{eqnarray}
where $G$ is the gravitational constant and $\ln[\Lambda]$ is a constant, called the Coulomb logarithm. In the Poisson equation and the $q$-integral, we use the inverse function $r(\phi,t)$ of the potential $\phi(r,t)$, as introduced in \citep{Henon_1961}, since the the potential monotonically decreases with $r$ at time $t$ in a state of the Virial equilibrium during the slow relaxation (Section \ref{subsec:Conditions}). \citet{Henon_1961} applied a re-normalization group method to the OAFP model and found that the model is invariant under the change of \emph{two parameters}. The method implies that the OAFP model needs more boundary or physical conditions to be self-similar. For this, \citet{Henon_1961} used the relation between the total mass and the cluster radius. However, against this result, the present section shows that the model must intrinsically evolve self-similarly without assigning any extra physical conditions.

\subsection{Minimizing the number of physical quantities at the equation-level}

We use $q$ as the argument of the functions $f$ and $\epsilon$, rather than using $\epsilon$ as the arguments of $f$ and $q$ since $q$ is the constant of integral both on the relaxation and dynamical time scales \citep{Spitzer_1988}. The resulting expressions have the maximum value of $q$ explicitly, which allows us to specify the size of the cluster to infinity, as follows,
\begin{eqnarray}
	\frac{\partial\, f}{\partial\, t}&=&16\pi^{2}G^{2}m^2N\ln[\Lambda]\nonumber\\
	&\,&\quad\times\frac{\partial\, }{\partial\, q}\left[f(q,t)\int^{q}_{0}f\left(q',t\right)\text{d}q'+\frac{\partial\, f}{\partial\, q}\frac{1}{\frac{\partial\, \epsilon}{\partial\, q}}\left(\int_{0}^{q}f(q',t)q'\frac{\partial\, \epsilon}{\partial\, q}\text{d}q'+q\int_{q}^{\infty}f(q',t)\frac{\partial\, \epsilon}{\partial\, q'}\text{d}q'\right)\right],\\
	0&=&r(\phi,t)\,\frac{\partial^{2}\, r}{\partial\, \phi^2}-2\left(\frac{\partial\, r}{\partial\, \phi}\right)^{2}+16\pi^{2}GMN\,r(\phi,t)\left(\frac{\partial\, r}{\partial\, \phi}\right)^{3}\int^{\infty}_{\kappa(\phi)}\sqrt{2(\epsilon(q',t)-\phi)}f(q',t)\frac{\partial\, \epsilon}{\partial\, q'}\text{d}q',\\
	q&=&\frac{1}{3}\int_{\phi(0,t)}^{\epsilon(q,t)}\left[2(\epsilon(q,t)-\phi')\right]^{3/2}r(\phi,t)^{2}\frac{\partial\, r}{\partial\, \phi'}\text{d}\phi',
\end{eqnarray}
where
\begin{equation}
	\kappa(\phi)=\frac{1}{3}\int_{\phi(0,t)}^{\phi}\left[2\left(\phi-\phi'\right)\right]^{3/2}r(\phi')^{2}\frac{\partial r}{\partial \phi'}\text{d}\phi'.
\end{equation}
We simplify the form of the OAFP model to reduce the number of the physical quantities based on two ideas. First, as mentioned in \citep{Henriksen_2015}, the general form of the Buckingham's Pi theorem may  include the translational invariance. The OAFP model is invariant only under the translational transformation of time $t$ . Second, we can define new physical quantities to reduce the number of governing parameters repeatedly appearing in the OAFP model. We use the fact that dimension equation of the OAFP equation reduces to $\left[\frac{1}{t}\right]=\left[m^2 G^2 N \ln[\Lambda] f\right]$. To encompass the aspect of translational invariance and avoid the repeated expressions, we introduce
\begin{eqnarray}
	s&\equiv&mG\ln[\Lambda](t+t_{T}),\\
	f^{*}(q,s)&\equiv&16\pi^{2}mGN\,f\left(q,\frac{s}{mG\ln[\Lambda]}-t_\text{T}\right),\\
	r^{*}(\phi,s)&\equiv&r\left(\phi,\frac{s}{mG\ln[\Lambda]}-t_\text{T}\right),\\
	\epsilon^{*}(q,s)&\equiv&\epsilon\left(q,\frac{s}{mG\ln[\Lambda]}-t_\text{T}\right),\\
	\epsilon_\text{c}(s)&\equiv&-\phi\left(0,\frac{s}{mG\ln[\Lambda]}-t_\text{T}\right),
\end{eqnarray}
where $t_\text{T}$ is constant. The infinite OAFP model with the new variables reads
\begin{eqnarray}
	\frac{\partial\, f^{*}}{\partial\, s}&=&\frac{\partial\, }{\partial\, q}\left[f^{*}(q,s)\int^{q}_{0}f^{*}\left(q',t\right)\text{d}q'+\frac{\partial\, f^{*}}{\partial\, q}\frac{1}{\frac{\partial\, \epsilon^{*}}{\partial\, q}}\left(\int_{0}^{q}f^{*}(q',s)q'\frac{\partial\, \epsilon^{*}}{\partial\, q}\text{d}q'+q\int_{q}^{\infty}f^{*}(q',s)\frac{\partial\, \epsilon^{*}}{\partial\, q'}\text{d}q'\right)\right],\\
	0&=&r^{*}\,\frac{\partial^{2}\, r^{*}}{\partial\, \phi^2}+\left(\frac{\partial\, r^{*}}{\partial\, \phi}\right)^{2}+r^{*}\left(\frac{\partial\, r^{*}}{\partial\, \phi}\right)^{3}\int^{\infty}_{\kappa(\phi)}\sqrt{2(\epsilon^{*}(q',s)-\phi)}f^{*}(q',s)\frac{\partial\, \epsilon^{*}}{\partial\, q'}\text{d}q',\\
	q&=&\frac{1}{3}\int_{-\epsilon_\text{c}(s)}^{\epsilon^{*}(q,s)}\left[2(\epsilon^{*}(q,s)-\phi')\right]^{3/2}r^{*}(\phi,s)^{2}\frac{\partial\, r^{*}}{\partial\, \phi'}\text{d}\phi'.
\end{eqnarray}

\subsection{Applying the Buckingham's Pi theorem}\label{sec3.2:Inf_Pi_theore}
When the total stellar number and energy of the OAFP model are indefinite, or infinitely large, there is no specific values of the physical quantities to characterize the infinite OAFP model. The physical quantities hence read
\begin{eqnarray}
	f^{*}(q,s)&=&F^{*}(q,s,\epsilon_\text{c}(s)),\\
	r^{*}(\phi,s)&=&R^{*}(\phi,s,\epsilon_\text{c}(s)),\\
	\epsilon^{*}(q,s)&=&E^{*}(q,s,\epsilon_\text{c}(s)).
\end{eqnarray}
These quantities are related to the governing parameters $\{q,s,\phi,\epsilon_\text{c}(s)\}$ through the physical relations $\{F^{*},R^{*},E^{*}\}$. The dimensions of the parameters are
\begin{eqnarray}
	\left[q\right]&=&\frac{L^6}{T^3},\qquad \left[s\right]=\frac{L^3}{T},\qquad \left[\epsilon_\text{c}(s)\right]=\left[\phi\right]=\left[\epsilon^{*}\right]=\frac{L^2}{T^2},\\
	\left[f^{*}\right]&=&\frac{T}{L^3},\qquad \left[r^{*}\right]=L,
\end{eqnarray}
where $L$ and $T$ represents the dimensions of length and time. Accordingly, the total number of the governing parameters for each physical relation is $n=3$, and that of the parameters with independent dimensions is $k=2$. The Buckingham's Pi theorem states each of $\{F^{*},R^{*},E^{*}\}$ depends only on ($m=$) 1 dimensionless parameter, as follows
\begin{eqnarray}
	\frac{f^{*}(q,s)}{s^{-1}}&=&F\left(\frac{q}{s^{3/2}\epsilon_\text{c}(s)^{3/4}}\right),\\
	\frac{\epsilon^{*}(q,s)}{\epsilon_\text{c}(s)}&=&E\left(\frac{q}{s^{3/2}\epsilon_\text{c}(s)^{3/4}}\right),\\
	\frac{r^{*}(\phi,s)}{s^{1/2}\epsilon_\text{c}(s)^{-1/4}}&=&R\left(\frac{\phi}{\epsilon_\text{c}(s)}\right),
\end{eqnarray}
where $\{F,E,R\}$ are functions. We may rewrite the final expression in terms of the original physical quantities
\begin{eqnarray}
	f(q,t)&=&\frac{1}{16\pi^{2}m^{2}G^{2}N(t+t_{T})\ln[\Lambda]}\,F\left(\frac{q}{([t+t_{T}]Gm\ln[\Lambda])^{3/2}\epsilon_\text{c}(s)^{3/4}}\right),\\
	\epsilon(q,t)&=&\epsilon_\text{c}(s)\,E\left(\frac{q}{([t+t_{T}]Gm\ln[\Lambda])^{3/2}\epsilon_\text{c}(s)^{3/4}}\right),\\
	r(\phi,t)&=&\frac{\sqrt{(t+t_{T})Gm\ln[\Lambda]}}{\epsilon_\text{c}(s)^{1/4}}R\left(\frac{\phi}{\epsilon_\text{c}(s)}\right).
\end{eqnarray}
	 
\subsection{Self-similarity of different situations in the infinite OAFP model}

The most intriguing finding in Section \ref{sec3.2:Inf_Pi_theore} is that the self-similarity of the infinite OAFP model is its intrinsic property and the outcome of the principle of covariance only. The self-similarity does not depend on neither the explicit form of $\epsilon_\text{c}(s)$ nor the value of $t_{T}$. This means that the property could appear in a broad class of the time-evolution of the model. There are four possible paths depending on $t_{T}$.
\begin{enumerate}
	\item  $t_\text{T}>0$ and $t>0$: core-expanding phase starting from a finite initial density
	\item  $t_\text{T}=0$ and $t>0$: core-expanding phase starting from an infinite initial density
	\item  $t_\text{T}=-t_\text{cc}<0$ and $0<t<t_\text{cc}$: (pre-collapse stage) core-collapsing phase until core-collapse time $t_\text{cc}$
	\item  $t_\text{T}=-t_\text{cc}<0$ and $t_\text{cc}<t$: (post-collapse stage) core-expanding phase after core-collapse time $t_\text{cc}$
\end{enumerate}
For each path, the explicit form of $\epsilon_\text{c}(s)$ must be determined with proper boundary conditions as done in \citep{Henon_1961} for the pre-collapse stage. When different possible solutions exist like above, a liner stability analysis generally help us check the validity of each. 

The rest of the present work focuses on the pre-collapse stage only. In this case, we retrieve the forms of self-similar variable assumed in the previous works \citep{Henon_1961,Heggie_1988},
\begin{equation}
	f(q,t)=f_\text{c}(t)\,F\left(\frac{q}{q_\text{c}(t)}\right),\qquad
	\epsilon(q,t)=\epsilon_\text{c}(t)\,E\left(\frac{q}{q_\text{c}(t)}\right),\qquad
	r(\phi,t)=r_\text{c}(t)\,R\left(\frac{\phi}{\epsilon_\text{c}(t)}\right),
\end{equation}
where the only the difference from the present works is constant factors in $q_\text{c}(t)$, $f_\text{c}(t)$, and $r_\text{c}(t)$.

	\section{Dimensional Analysis on a finite OAFP model}\label{sec4:Fin_OAFP}

Imagine an isotropic finite star cluster in a static tidal field due to its host galaxy. The corresponding finite OAFP system reads
\begin{eqnarray}
	\frac{\partial\, f^{*}}{\partial\, s}&=&\frac{\partial\, }{\partial\, q}\left[f^{*}(q,s)\int^{q}_{0}f^{*}\left(q',t\right)\text{d}q'+\frac{\partial\, f^{*}}{\partial\, q}\frac{1}{\frac{\partial\, \epsilon^{*}}{\partial\, q}}\left(\int_{0}^{q}f^{*}(q',s)q'\frac{\partial\, \epsilon^{*}}{\partial\, q}\text{d}q'+q\int_{q}^{q_\text{M}}f^{*}(q',s)\frac{\partial\, \epsilon^{*}}{\partial\, q'}\text{d}q'\right)\right],\label{Eq.finite_OAFP}\\
	0&=&r^{*}\,\frac{\partial^{2}\, r^{*}}{\partial\, \phi^2}-2\left(\frac{\partial\, r^{*}}{\partial\, \phi}\right)^{2}+r^{*}\left(\frac{\partial\, r^{*}}{\partial\, \phi}\right)^{3}\int^{q_\text{M}}_{\kappa(\phi)}\sqrt{2(\epsilon^{*}(q',s)-\phi)}f^{*}(q',s)\frac{\partial\, \epsilon}{\partial\, q'}\text{d}q',\\
	q&=&\frac{1}{3}\int_{-\epsilon_\text{c}(s)}^{\epsilon^{*}(q,s)}\left[2(\epsilon^{*}(q,s)-\phi')\right]^{3/2}r^{*}(\phi,s)^{2}\frac{\partial\, r^{*}}{\partial\, \phi'}\text{d}\phi',
\end{eqnarray} 
with the conservation laws for the total stellar number and energy
\begin{eqnarray}
	\int_{0}^{q_\text{M}}f^{*}\left(q',t\right)\text{d}q'&=&\int_{0}^{q_\text{M}}f^{*}_\text{o}\left(q'\right)\text{d}q',\label{Eq.conser_N}\\
	\int_{0}^{q_\text{M}}f^{*}\left(q',t\right)\,\left[\frac{1}{2}\epsilon^{*}\left(q',t\right)+\frac{3q'}{4}\frac{\partial \epsilon^{*}}{\partial q'}\right]\text{d}q'&=&\int_{0}^{q_\text{M}}f^{*}_\text{o}\left(q'\right)\,\left[\frac{1}{2}\epsilon^{*}_\text{o}\left(q'\right)+\frac{3q'}{4}\frac{\partial \epsilon^{*}_\text{o}}{\partial q'}\right]\text{d}q',
\end{eqnarray}
where $q_\text{M}$ is the maximum of $q$. The initial conditions for $f^{*}\left(q,s\right)$ and $\epsilon^{*}\left(q,s\right)$ are
\begin{eqnarray}
	f^{*}_\text{o}\left(q\right)&\equiv&f^{*}\left(q,s=s_\text{o}\right),\\
	\epsilon^{*}_\text{o}\left(q\right)&\equiv&\epsilon^{*}\left(q,s=s_\text{o}\right),
\end{eqnarray}
where $s_\text{o}$ is the value of $s$ at $t=0$.

\subsection{Applying the Buckingham's Pi theorem}
The physical quantities are, in terms of the governing parameters,
\begin{eqnarray}
	f^{*}(q,s)&=&F^{*}\left(q,s,\epsilon_\text{c}(s),q_\text{M},f^{*}_\text{o}(q)\right),\\
	r^{*}(\phi,s)&=&R^{*}\left(\phi,s,\epsilon_\text{c}(s),q_\text{M},r^{*}_\text{o}(\phi)\right),\\
	\epsilon^{*}(q,s)&=&E^{*}\left(q,s,\epsilon_\text{c}(s),q_\text{M},\epsilon^{*}_\text{o}(q)\right),
\end{eqnarray}
where the initial condition for $r^{*}(\phi,s)$ is
\begin{equation}
	r^{*}_\text{o}\left(\phi\right)\equiv r^{*}\left(\phi,s=s_\text{o}\right).
\end{equation}
As $q_\text{M}$ explicitly appears in the Poisson equation for $r^{*}(\phi,s)$, we use $q_\text{M}$ as a governing parameter of $R^{*}$ rather than its maximum value $r^{*}(0,s)$.

The dimensions of the parameters are
\begin{eqnarray}
	\left[q\right]&=&\left[q_\text{M}\right]=\frac{L^6}{T^3},\qquad \left[s\right]=\frac{L^3}{T},\qquad \left[\epsilon_\text{c}(s)\right]=\left[\phi\right]=\left[\epsilon^{*}\right]=\left[\epsilon^{*}_\text{o}\right]=\frac{L^2}{T^2},\\
	\left[f^{*}\right]&=&\left[f^{*}_\text{o}\right]=\frac{T}{L^3},\qquad \left[r^{*}\right]=\left[r^{*}_\text{o}\right]=L.
\end{eqnarray}
Accordingly, $n=5$ and $k=2$. The Buckingham's Pi theorem reduces the number of dimensional governing parameters into $(m=)3$ dimensionless ones, as follows
\begin{eqnarray}
	\frac{f^{*}(q,s)}{s^{-1}}&=&F^{*}\left(\frac{q}{q_\text{c}(s)},\frac{f_\text{o}^{*}(q)}{s^{-1}},\frac{q_\text{M}}{q_\text{c}(s)}\right),\\
	\frac{\epsilon^{*}(q,s)}{\epsilon_\text{c}(s)}&=&E^{*}\left(\frac{q}{q_\text{c}(s)},\frac{\epsilon^{*}_\text{o}(q)}{\epsilon_\text{c}(s)},\frac{q_\text{M}}{q_\text{c}(s)}\right),\\
	\frac{r^{*}(\phi,s)}{r_\text{c}(s)}&=&R^{*}\left(\frac{\phi}{\epsilon_\text{c}},\frac{r^{*}_\text{o}(\phi)}{r_\text{c}(s)},\frac{q_\text{M}}{q_\text{c}(s)}\right).
\end{eqnarray}
We limit our focus into complete similarity (Section\ref{subsec:Conditions}). As $s\to 0$, the boundary conditions to be assigned on the physical relations at $q=q_\text{M}$ and $\phi=-\epsilon_{c}(s)$, are
\begin{equation}
	F^{*}\left(\frac{q_\text{M}}{q_\text{c}(s)},0,\frac{q_\text{M}}{q_\text{c}(s)}\right)=0,\quad
	E^{*}\left(\frac{q_\text{M}}{q_\text{c}(s)},0,\frac{q_\text{M}}{q_\text{c}(s)}\right)=0,\quad
	R^{*}\left(-1,0,\frac{q_\text{M}}{q_\text{c}(s)}\right)=0,	
\end{equation}
and, at $q=0$ and $\phi=0$, are
\begin{equation}
	0<F^{*}\left(0,\frac{f_\text{o}^{*}(0)}{s^{-1}},\frac{q_\text{M}}{q_\text{c}(s)}\right)<\infty,\quad
	0<E^{*}\left(0,\frac{\epsilon_\text{o}^{*}(0)}{\epsilon_\text{c}(s)},\frac{q_\text{M}}{q_\text{c}(s)}\right)<\infty,\quad
	0<R^{*}\left(0,\frac{r^{*}_\text{o}(0)}{r_\text{c}(s)},\frac{q_\text{M}}{q_\text{c}(s)}\right)<\infty.\label{BC:F*}
\end{equation}
With these boundary conditions, the model's central density is always finite (except for $s=0$) and zero at $q=q_\text{M}$ because of the monotonicity of the functions (Section \ref{subsec:Conditions}). Hence, we have the asymptotic conditions for dimensionless parameters as $s\to0$ are
\begin{eqnarray}
	0<F\left(\frac{q}{q_\text{c}(s)}\right)&\equiv&F^{*}\left(\frac{q}{q_\text{c}(s)},0,\infty\right)<\infty,\\
	0<E\left(\frac{q}{q_\text{c}(s)}\right)&\equiv&E^{*}\left(\frac{q}{q_\text{c}(s)},\infty,\infty\right)<\infty,\\
	0<R\left(\frac{\phi}{\epsilon_\text{c}(s)}\right)&\equiv&R^{*}\left(\frac{\phi}{\epsilon_\text{c}(s)},\infty,\infty\right)<\infty.
\end{eqnarray}   
     
\subsection{Power-law profile in the finite OAFP model}

With the boundary and asymptotic conditions, we evaluate the conservation laws. The conservation of total number \eqref{Eq.conser_N} is, in terms of the dimensionless variables, 
\begin{equation}
	\int_{0}^{q_\text{M}}\frac{1}{s}F^{*}\left(\frac{q'}{q_\text{c}(s)},\frac{f_\text{o}^{*}(q)}{s^{-1}},\frac{q_\text{M}}{q_\text{c}(s)}\right)\text{d}q'=\int_{0}^{q_\text{M}}f^{*}_\text{o}\left(q'\right)\text{d}q'=\text{const.}
\end{equation}
Based on Section \ref{subsec:Conditions}, the outer halo is collisionless and the distribution function is stationary near $q=q_\text{M}$. We exclude the contribution to the conservation law from large phase-space volumes, as follows,
\begin{equation}
	\int_{0}^{q_\text{s}}\frac{1}{s}F^{*}\left(\frac{q'}{q_\text{c}(s)},\frac{f_\text{o}^{*}(q)}{s^{-1}},\frac{q_\text{M}}{q_\text{c}(s)}\right)\text{d}q'=\int_{0}^{q_\text{s}}f^{*}_\text{o}\left(q'\right)\text{d}q'=\text{const.},
\end{equation}
where $0\ll q_\text{s}<q_\text{M}$. We introduce the constant $q_\text{s}$ only for convenience to characterize the boundary between the inner and outer halos. There is no way to strictly determine the value as it depends on the initial condition and the stage of evolution. Introducing new variable $Q=q/q_\text{c}(s)$, we have
\begin{equation}
	\int_{0}^{q_\text{s}/q_\text{c}(s)}F^{*}\left(Q',\frac{f_\text{o}^{*}(Qq_\text{c}(s))}{s^{-1}},\frac{q_\text{M}}{q_\text{c}(s)}\right)\text{d}Q'=\frac{s}{q_\text{c}(s)}\int_{0}^{q_\text{s}}f^{*}_\text{o}\left(q'\right)\text{d}q'\propto \frac{s}{q_\text{c}(s)},\label{Eq.power_law_test}
\end{equation}
The explicit form of $\epsilon_\text{c}(s)$ and $q_\text{c}(s)$ are unknown so far. We hence consider four possible cases to determine the time-dependence of $q_\text{c}(s)$. For small $s$ or the limit $s\to 0$, (i) $q_\text{c}(s)$ is finite, (ii) $q_\text{c}(s) \to \infty$, (iii) $q_\text{c}(s)\to0$ and the leading-order of $F$ is not a power of $s$, and (iv) $q_\text{c}(s)\to0$ and the leading-order of $F$ is a power of $s$.
 
Case (i): Imagine that $s\to 0$ and $q_\text{c}(s)$ is constant. Equation \eqref{Eq.power_law_test} then reduces to
\begin{equation}
	\int_{0}^{q_\text{s}/q_\text{c}(s)}F^{*}\left(Q',0,\frac{q_\text{M}}{q_\text{c}(s)}\right)\text{d}Q'\propto\frac{1}{s}\to\infty.
\end{equation} 
As the integral must be finite for finite $q_\text{c}(s)$, this case is inconsistent as $s\to0$. 

Case (ii): The relation $F^{*}$ has the following form in the limit $q_\text{c}(s)\to\infty$
\begin{equation}
	F^{*}(0,0,0)\propto s\to 0,
\end{equation} 
which is also inconsistent as $F^{*}(0,0,0)$ describes the core state and should be finite and non-zero. 

Case (iii): For  a very small $s$ and $q_\text{c}(s)$, we have the approximate of equation \eqref{Eq.power_law_test}
\begin{equation}
	\int_{0}^{q_\text{s}/q_\text{c}(s)}F^{*}\left(Q',0,\infty\right)\text{d}Q'\propto \frac{s}{q_\text{c}(s)}.
\end{equation}
Even for a small $q_\text{c}(s)$, the integral needs to be finite, as it is simply a partial integral of the distribution function for a finite system. Hence, $q_\text{c}(s)\propto s$. Other relaxation or dynamical effects are not expected in the limits of $s\to0$ and $q_\text{c}(s)\to 0$, so we have
\begin{equation}
	\int_{0}^{\infty}F^{*}\left(Q',0,\infty\right)\text{d}Q'=\text{const.}
\end{equation}
This relation, however, is inconsistent. With the same limits, the finite OAFP system exactly reduces to the same mathematical forms as the infinite OAFP system. According to the result of Section \ref{sec3:Inf_OAFP}, the OAFP model is always self-similar \emph{if} the total number and energy are indefinite or infinite. The above conservation law does not satisfy the condition. 

Case (iv): For a very small $s$ and $q_\text{c}(s)$, we obtain the same approximate as the case (iii), but with the leading-order of $F^{*}$ proportional to a power-law profile $\sim Q^{-\gamma}$, where $\gamma$ is a real number,
\begin{equation}
	\int_{0}^{q_\text{s}/q_\text{c}(s)}F^{*}\left(Q',0,\infty\right)\text{d}Q'\sim \int_{0}^{q_\text{s}/q_\text{c}(s)} {Q'}^{-\gamma}\text{d}Q' \propto \frac{s}{q_\text{c}(s)}.
\end{equation}
The powers of $s$ can balanced out at the equation level, in the limit $s\to 0$, if
\begin{eqnarray}
	&\,&q_\text{c}(s)\propto s^{1/\gamma},\\
	&\,&\epsilon_\text{c}(s)\propto s^{-2+4/(3\gamma)},
\end{eqnarray}
where $7/9<\gamma<2/3$ is the necessary condition for an infinite self-gravitating system to have a power-law profile, as spherical polytropes of $m=5$ to $m=\infty$ are infinite in size \citep{Chandra_1939}. Within the range of $\gamma$, the total number and energy always diverge, in the limit $s\to\infty$,

\begin{eqnarray}
	&\,&\int_{0}^{\infty}F^{*}\left(Q',0,\infty\right)\text{d}Q'\propto s^{1-1/\gamma}\to\infty,\\
	&\,&\int_{0}^{\infty}F^{*}\left(Q',0,\infty\right)\,\left[\frac{1}{2}E^{*}\left(Q',0,\infty\right)+\frac{3Q'}{4}\frac{\partial E^{*}}{\partial Q'}\right]\text{d}Q'\propto s^{3-7/(3\gamma)}\to\infty,
\end{eqnarray}
where $E^*$ approximates to a power of $Q$ as $s\to0$ since, if it is not a power, the same inconsistent situation occurs as case (iii). Accordingly, only case (iv) is consistent with the conservation laws and the result of Section \ref{sec3:Inf_OAFP}, obtained from the principle of covariance. Once $F^{*}$ and $E^{*}$ approximate to powers of $Q$ as $s\to0$, the corresponding Poisson equation is the Lane-Emden equation \citep{Chandra_1939} for a singular spherical polytropes of $m>5$. This means that $R^{*}$ approximates to a power of $-\phi$. Hence, the density $\rho(r,t)$ of the finite OAFP model must have a power-law profile; $\rho(r,t)\propto r^{-\alpha}$ ($2<\alpha<2.5$) at $q\lesssim q_\text{s}$, as $s\to0$.

We lastly mention where the inner halo is in a galactic globular cluster. There is no strict definition to locate the boundary of the inner halo from the core, in the same way as the outer-inner halo boundary. If we use the conventional definition, the inner halo can be limited to $q_\text{c}(s)\lesssim q\lesssim q_\text{s}$. At the order of $q=q_\text{c}(s)$, the slope of the power-law profile needs to be calm so that $F^{*}$ is finite at the core ($q\lesssim q_\text{c}(s)$) to satisfy the boundary condition \eqref{BC:F*}. This condition is essential. Without the condition, an infinite density still satisfies the conservation laws. The contribution of the singular power-law profile to the integrals in the conservation laws is subtle as the phase-space volume $(\sim Q)$ decays faster than the distribution function $(\sim Q^{-\gamma})$ blows up, as $Q\to0$. \citet{Heggie_1988} numerically confirmed the validity of the power-law structure, by assuming that the cluster has power-law at $q_\text{c}(s)\gtrsim q$ and comparing to the result of \citep{Cohn_1980}. Hence, the finite OAFP model must reveal a power-law profile at $q_\text{c}(s)\lesssim q\lesssim q_\text{s}$, or the inner halo, as $s\to 0$.
	\section{Conclusion}\label{sec5:Conclusion}

To elucidate the origin of the power-law profile in the inner halo of a core-collapsing galactic globular cluster, the region is conventionally assumed to hold a self-similarity and a stationary state based on local analysis. The present paper applied dimensional analysis to a finite and an infinite OAFP model for an isotropic star cluster of equal masses. We showed that the power-law profile originates from the principle of covariance and the conservation laws.

Realistic galactic globular clusters, however, have richer structures and dynamical events, such as mass distribution and mass loss, which increases the number of dimensionless governing parameters. Accordingly, it is essential to extend the present work to a dimensional analysis on more realistic clusters.

From the kinetic-theoretical point of view, it is also important to extend the present work to inhomogeneous Landan equation \citep{Polyachenko_1982} and inhomogeneous Balescu-Lenard equation \citep{Heyvaerts_2010,Chavanis_2012}. The equations include more realistic effects in the collision terms than OAFP equation, such as inhomogeneity and gravitational amplification. Their numerical computations have been anticipated for globular-cluster studies, but not been carried out because of their complex mathematical structures. Hence, it is where a dimensional analysis comes in and helps to predict possible numerical results and mathematical structures.  
	
	
	
	\bibliographystyle{elsarticle-harv} 
	\bibliography{science}
	

\end{document}